\documentclass[a4paper,11pt]{article}
\usepackage{pos}

\newcommand{\pTf}{p_{\scriptscriptstyle T}}
\newcommand{\pTi}{p_{\scriptscriptstyle T}^{\rm \tiny initial}}

\newcommand{\be}{\begin{equation}}
\newcommand{\ee}{\end{equation}}

\title{Jet tomography in hot QCD medium with deep learning}
\author*[a]{Yi-Lun Du}
\author[b]{Daniel Pablos}
\author[a]{Konrad Tywoniuk}

\affiliation[a]{Department of Physics and Technology, University of Bergen,\\ Postboks 7803, 5020 Bergen, Norway}

\affiliation[b]{INFN, Sezione di Torino, via Pietro Giuria 1, \\ I-10125 Torino, Italy}

\emailAdd{yilun.du@uib.no}
\emailAdd{daniel.pablos.alfonso@to.infn.it}
\emailAdd{Konrad.Tywoniuk@uib.no}

\abstract{With deep learning techniques, the degree of modification of energetic jets that traversed hot QCD medium can be identified on a jet-by-jet basis. Due to the strong correlations between the degree of jet modification and its traversed length in the medium, we demonstrate the power of our novel method to locate the creation point of a dijet pair in the nuclear overlap region. In particular, jet properties, such as jet width and orientation can serve as additional handles to locate the creation points to a higher level of precision, which constitutes a significant development towards the long-standing goal of using jets as tomographic probes of the quark-gluon plasma.}

\FullConference{%
  *** The European Physical Society Conference on High Energy Physics (EPS-HEP2021), ***\\
  *** 26-30 July 2021 ***\\
  *** Online conference, jointly organized by Universität Hamburg and the research center DESY ***
}

\begin{document}
\maketitle

\section{Introduction}
Jets are narrow cones of hadrons that are produced in hard QCD processes in high-energy particle collisions. 
% \cite{Salam:2009jx,Larkoski:2017jix,Marzani:2019hun}. 
In heavy-ion collisions, they form concurrently with the creation of hot and dense QCD matter, known as the quark-gluon plasma (QGP), which
%This new state of matter, which is believed to have filled the Universe during the first microseconds after the Big Bang, 
behaves like a nearly perfect liquid.
%the most perfect liquid ever measured in Nature 
% \cite{ackermann2001elliptic,aamodt2010elliptic,aamodt2011higher}.
When passing through this medium, partonic jet modes will experience momentum diffusion and energy loss. The latter phenomenon, known as jet quenching, occurs by the radiation of soft particles towards large angles \cite{dEnterria:2009xfs,Majumder:2010qh,Mehtar-Tani:2013pia,Blaizot:2015lma}. Unique properties of the medium are contained in the detailed modifications of these hard probes, which turn them into useful tools, and therefore tremendous theoretical and experimental effort is being devoted on them. 
% \cite{Abelev:2013kqa,Adam:2015ewa,aad2015measurements,Aaboud:2017eww,aaboud2019measurement,Acharya:2019jyg}.
%In practice, a jet is defined through a reconstruction algorithm with an associated parameter $R$ that can be interpreted as an angular size. The momenta of the collection of particles that belong to the jet are recombined, usually using sequential addition of four-momenta, to determine the kinematics of the jet itself.
%Jets possess an internal structure, a recognizable organization pattern of the particles within, which is well understood in vacuum~\cite{Salam:2009jx,Larkoski:2017jix,Marzani:2019hun}. 
%For instance, the individual splittings that make up the tree-structure of a jet can also be associated with a timescale \cite{Andrews:2018jcm}, given by the quantum-mechanical formation time \cite{Dokshitzer:1991wu}.
%In properly defined infrared and collinear safe observables, non-perturbative effects are strongly mitigated allowing for a direct access to the perturbative structure of jets.

Using jets as differential probes of the evolution of the QGP created in heavy-ion collisions, aka \emph{jet tomography}, is a long-term quest \cite{Wang:2002ri,Renk:2006qg,Zhang:2007ja,Zhang:2009rn,He:2020iow,PhysRevLett.127.082301}. The medium-induced modifications of jets follow from the local properties of the medium along the jet trajectory. The capability to unambiguously capture the details of these interactions, for each individual jet, would lead to unprecedented precision in determining local properties of the fluid, including flow \cite{Armesto:2004vz,Sadofyev:2021ohn}, path-length dependence of jet modifications \cite{Betz:2014cza} and improved possibility of observing deconfined quasi-particles as degrees of freedom in the QGP \cite{DEramo:2018eoy,Barata:2020rdn,Harris:2020ijy}. 
% Nonetheless, tomographic analyses on the level of inclusive jet populations have been hindered by intrinsic biases that accentuate samples experiencing small modifications over samples that are strongly affected~\cite{Baier:2001yt}. Such biases arise due to the steeply falling spectrum of the jet initiator transverse momenta and strongly distort the magnitude of medium effects, e.g. the in-medium path-length distribution of surviving jets. 

%Jets in heavy-ion collisions have been proposed as tomographic probes of the QGP since the beginning of the field~\cite{Wang:2002ri,Renk:2006qg,Zhang:2007ja,Zhang:2009rn}. 
%One of the main ideas has consisted in contrasting the distribution of production points, and related distribution of in-medium path-lengths, of different hard probe samples, for instance inclusive jets (or hadrons) in dijet events versus semi-inclusive boson-jet (or boson-hadron) samples, on which the selection bias has a different effect due to the different associated jet production spectrum. More recently, a new technique called gradient jet tomography has been developed, which exploits the correlation between the original production point and the asymmetry in the  transverse momentum distribution, with respect to the jet direction, of partons in the cone caused by the spatial gradient of the jet quenching parameter $\widehat{q}$~\cite{He:2020iow}. 
%Getting access to this information is a key step towards future, more focused studies in which the modifications of quenched jets can provide detailed aspects of the properties of the QGP along its propagation path.

In this paper we employ deep learning techniques to estimate the energy loss suffered by a given reconstructed jet at $\pTf$ and cone size $R$ measured in a heavy-ion collision.
% that mitigates these bias effects and results in a better control of the path-length traversed by individual jets based on their level of modification. 
% Given a measured jet at $\pTf$ and cone size $R$, the procedure allows us to estimate with reasonable accuracy the transverse momentum $p_T^{\rm initial}$ 
% the jet would have had, had it not interacted with a medium, see \cite{Du:2020pmp} for further details on how to establish such a correspondence.
Having this extracted knowledge at hand allows for many interesting applications, such as revealing more pronounced substructure modifications of jets \cite{Du:2020pmp}, getting access to the genuine configuration profile of jets over the nuclear overlap region in the collision, both with respect to their creation points and orientations \cite{du2021jet} and using jets as tomographic probes of the QGP, which is the main focus of this proceedings. 
%It was previously used to reveal more pronounced substructure modifications of jets by constructing ratio observables for specific jet samples where one compares jets in proton-proton and heavy-ion collisions with the same \emph{initial} $p_T$ instead of comparing them at a fixed \emph{final} $p_T$, as conventionally done. 
%We refer to these two jet selections as \emph{initial energy selection} (IES) and \emph{final energy selection} (FES), respectively. 
% Here, we demonstrate the usefulness of our approach to tomographic applications in two concrete examples. 
% The first deals with reconstructing the true distribution of path-lengths that  jets experience, eliminating the effects of ``surface bias'' \cite{Dainese:2004te,Zhang:2007ja,Zhang:2009rn} and revealing the potential contributions to jet azimuthal anisotropy that do not stem from final-state interactions.
%The genuine nuclear density distributions that influence the hard process production point are in this way revealed. Moreover, possible initial-state effects that lead to anisotropy in the jet azimuthal orientation distribution could thereby be exposed. 

First, we demonstrate the power of our method to constrict the creation-point of a dijet pair over the nuclear overlap region by constraining the jet energy loss. Then, we present the improvement of the combination of the extraction of the lost energy with additional accessible knowledge about the jet width or the jet orientation with respect to the event plane of the collision. They allow to constrain the path length dependence separately for jets of specific width or jets traveling parallel and transverse to the event plane of the collisions, which refine the path to experimentally pinning down the original creation point of a dijet pair. These improvements contribute to the set of tools aimed at exploiting energetic jets as tomographic probes of the QGP.

\section{Jet energy loss with deep learning}
We first estimate, on a jet-by-jet basis, the amount of jet energy loss during the passage through a hot QCD medium, quantified through the variable $\chi = \pTf/\pTi$ within the hybrid strong/weak coupling model \cite{casalderrey2015erratum,Casalderrey-Solana:2015vaa,Casalderrey-Solana:2016jvj}.
% \be
% \label{eq:chi-definition}
% \chi \equiv \frac{\pTi}{p_T} \,,
% \ee
$p_T$ is the measurable transverse momentum of a given jet in the presence of a medium, and $\pTi$ is the transverse momentum of the \emph{same} jet would have had, had there been no medium. For further details on how to establish such a correspondence, see \cite{Du:2020pmp}. In the hybrid model, the vacuum evolution is factorized from the interactions with the medium. Other jet quenching models with this general picture should also allow for such a jet-by-jet correspondence \cite{Du:2020pmp,du2021jet}. 

% \begin{figure}[t!]
% \centering
% % \includegraphics[width=0.48\textwidth]{Prediction_performance_chijh_overlap.pdf}
% \includegraphics[width=0.6\textwidth]{Prediction_performance_chi_chip_overlap_300.pdf}
% \caption{Prediction performance of CNN. The green color represents the probability of predicted $\chi^p$ within the given true $\chi$ bin in the column-normalized 2-D histogram. The red line with error bar quantifies the average and standard deviation of the predicted $\chi^p$ within the given true $\chi$ bin.}
% \label{Prediction Performance}
% \end{figure}
Within the hybrid model, we generate around 250,000 jets at $\sqrt{s_{NN}}= 5.02$ TeV for PbPb collisions at 0-5\% centrality, which are reconstructed with FastJet \textcolor{black}{3.3.1} \cite{Cacciari:2011ma} using the anti-$k_T$ algorithm \cite{Cacciari:2008gp} with reconstruction parameter $R=0.4$, $|\eta|<2$ and measured $p_T>100$ GeV. 80\% of these samples are used for the training of our algorithm and the remaining 20\% are used validation. We train the convolutional neural network (CNN) to predict the energy loss ratio $\chi$ from the jet image in a supervised manner. The trained CNN can predict $\chi$ over a wide range with reasonable accuracy. For more details see Ref. \cite{Du:2020pmp}. 
% Fig.~\ref{Prediction Performance} shows the $\chi$ prediction performance. The joint distribution in green is column-normalized which represents the probability of predicted $\chi^p$ within the given true $\chi$ bin. The red line with error bar quantifies the average and standard deviation of the predicted $\chi^p$ within the given true $\chi$ bin. 

\section{Jet tomography}

\begin{figure}[t!]
\centering
\includegraphics[width=0.8\textwidth]{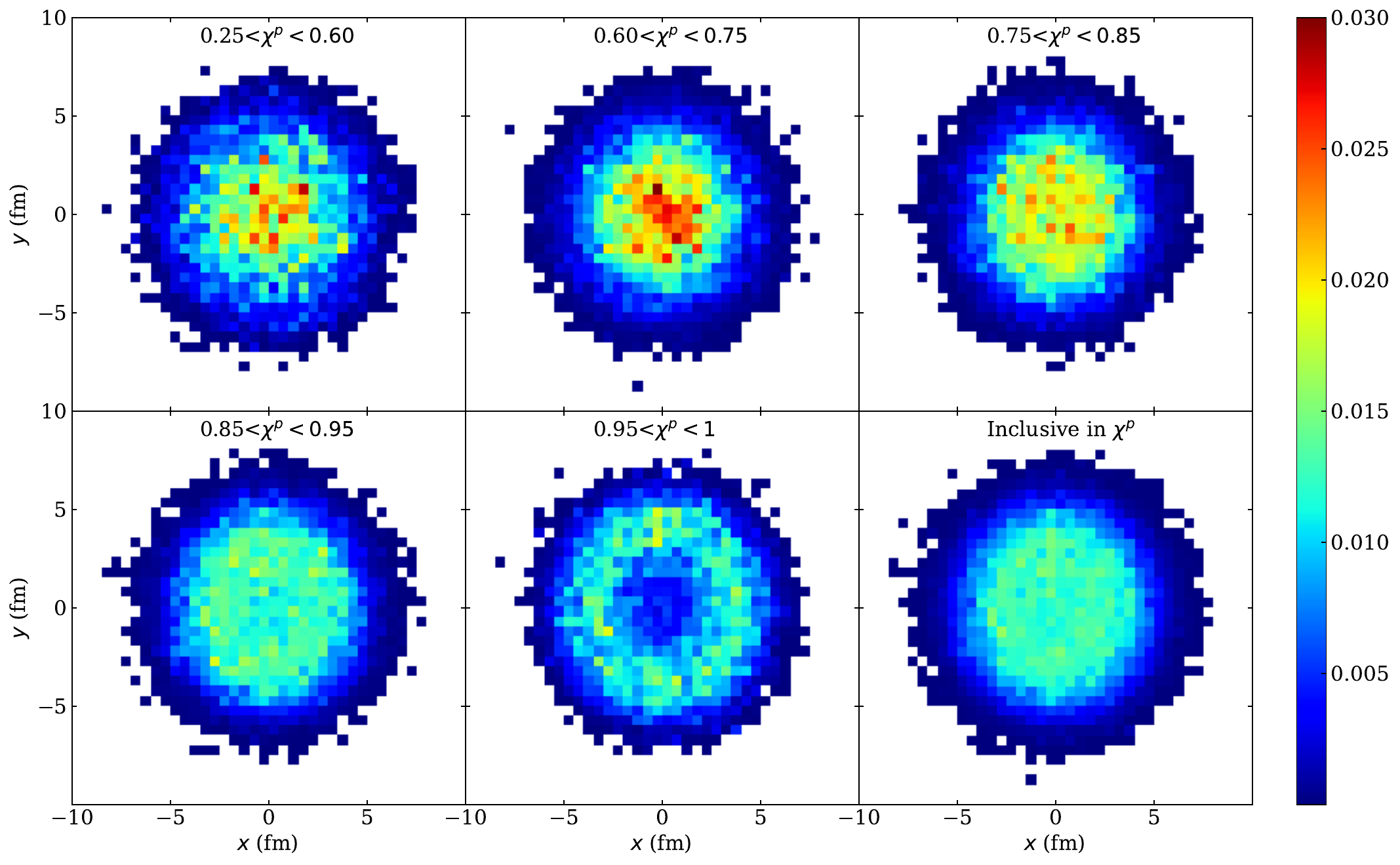}
\caption{Creation point distributions of jets in the transverse plane for different ranges of the predicted $\chi^p$.}
\label{creation}
\end{figure}

The traversed length by a given jet in QGP is strongly correlated to its energy loss and modifications. Specifically, by selecting jets suffering different amounts of energy loss $\chi$ extracted by our network, we are actually selecting jets traversing different lengths and consequently created at different positions in the transverse plane of the medium. This new capability paves the way to extract properties of QGP through jet tomographic applications with unprecedented precision.\footnote{Alternatively, the degree of medium induced jet modification can be reasonably estimated from the ratio of the jet $p_T$ over that of a recoiling colorless trigger boson, although their correlation is not tight due to sizeable out-of-cone radiation even in vacuum \cite{Zhang:2009rn}. Besides, a recent tomographic study based on spatial-temporal gradient of QGP also gives clue to the creation point of an energetic jet~\cite{He:2020iow}.}

To corroborate this statement, we can look at the jet creation-points distribution in the transverse plane $\lbrace x,y\rbrace$ sliced in $\chi$, which is shown in Fig.~\ref{creation}. It's worth pointing out, again, that the pair values of creation positions $\lbrace x,y\rbrace$ are not extracted by our algorithm, but taken directly from the Monte Carlo instead. One can see that jets of the little quenched class, $0.95<\chi^p<1$, were created within a ring structure at the periphery of the nuclear overlap collision region. The creation points move towards the centre of the initial geometry gradually as energy loss increases. It's worth mentioning that for the most quenched class, $0.25<\chi^p<0.6$, the distribution becomes more spread than that for the neighbour less quenched class, $0.6<\chi^p<0.75$. The reason can be attributed to that the most quenched jets have had to go through the largest traversed lengths, which correspond to jets that were created at the periphery and flew inwards, towards the center of the QGP. This observation inspires us to consider the jet orientation as an additional handle in a refined study. We will come back to this direction a bit later in the non-central collisions.

\begin{figure}[t!]
\centering
\includegraphics[width=0.8\textwidth]{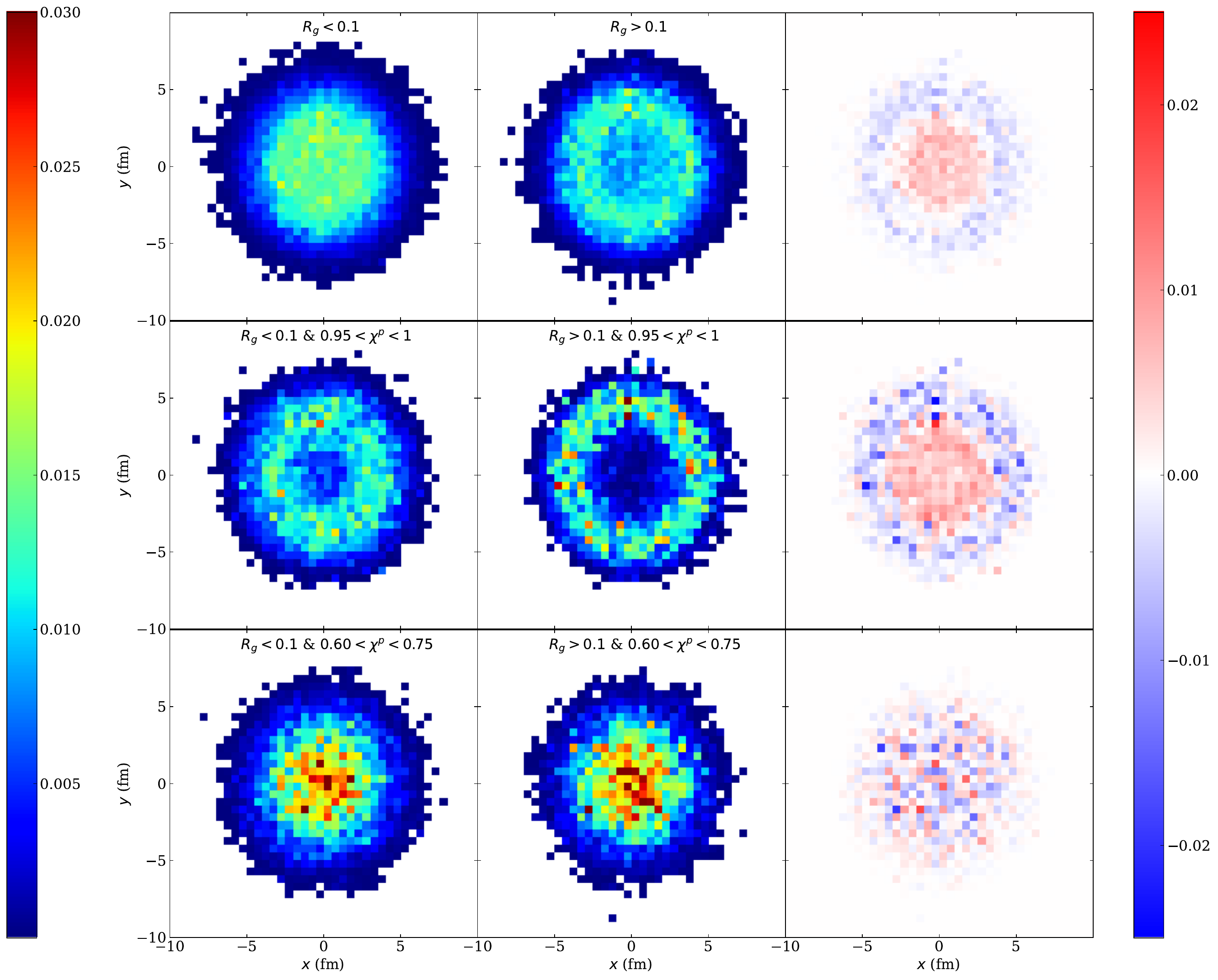}
\caption{Creation point distributions in the transverse plane for jets with measured $R_g<0.1$ (left column) versus $R_g>0.1$ (middle column), and their difference (right column). Jets are inclusive in $\chi$ (upper row), in little quenched class with $0.95<\chi^p<1$ (middle row) and very quenched class with $0.6<\chi^p<0.75$ (lower row).}
% \caption{The normalized distribution of the creation point in the transverse plane for jets with measured $R_g<0.1$ (left column) versus $R_g>0.1$ (middle column), and their difference (right column). Jets are required to have $p_T>100$ GeV and are inclusive in $\chi$ (upper row), in little quenched class with $0.95<\chi<1$ (middle row) and very quenched class with $0.6<\chi<0.75$ (lower row).}
\label{deltaRcreation}
\end{figure}

We have shown that with the extracted energy loss we have control of how long the jets have traversed within the QGP, which is an important step forward towards using jets as tomographic probes. One possible improvement of the above analysis could include more differential sample selections based on measurable jet properties, such as jet width $R_g$ within the Soft-Drop procedure~\cite{Larkoski:2014wba}, with parameters $z_{\textrm{cut}}=0.1$ and $\beta=0$. Given the dependence of energy loss on jet width, and its relation to selection bias, as discussed in Ref.~\cite{Du:2020pmp}, it is natural to expect that jets with different (initial) widths, and the same amount of the energy loss, will have traversed different lengths on average. To support this picture, we show in Fig.~\ref{deltaRcreation} the creation point distributions in the transverse plane for jets with a measured (this is, final) $R_g$ smaller or larger than 0.1. In the top row (inclusive in $\chi$), we see that narrower jets tend to pass the selection even if they were created deep inside the medium, while wider jets are pushed towards the surface in comparison. The little quenched jets, $0.95<\chi^p<1$, (middle row) agree with this picture as the more pronounced ring structure shows. For the very quenched jets, $0.6<\chi^p<0.75$, one can see the reverse of such ordering as shown in the lower row, which is similar as in Fig.~\ref{creation}, since to get such quenched, narrow jets tend to be created towards the periphery and flow inwards.

% Indeed, given what we know about the dependence of energy loss on the width of a jet, and its relation to selection bias, as discussed in Ref.~\cite{Du:2020pmp}, it is natural to expect that jets with different (initial) widths, and the same amount of the energy loss $\chi$, will on average have traversed different amounts of length. In Fig.~\ref{deltaRcreation} we show some results supporting this picture (using the true value of $\chi$ \Du{replace?}). In this figure we show the creation point distribution in the transverse plane, inclusive in $\chi$ (upper row), for jets with a measured (this is, final) $R_g$ smaller or larger than 0.1 (here again, $\chi$ could be estimated by the neural network while the production points are taken from the hybrid model). We see that narrower jets tend to pass the selection even if they were produced deep inside the medium, while wider jets are pushed towards the surface in comparison. The little quenched jets (middle row) agree with this picture as the more pronounced ring structure shows. Similar as discussed in Fig.~\ref{creation}, such ordering is reversed for the very quenched class shown in the lower row of Fig.~\ref{deltaRcreation}, since narrow jets that get such quenched tend to be produced towards the periphery and travel inwards.

\begin{figure}[t!]
\includegraphics[width=0.99\textwidth]{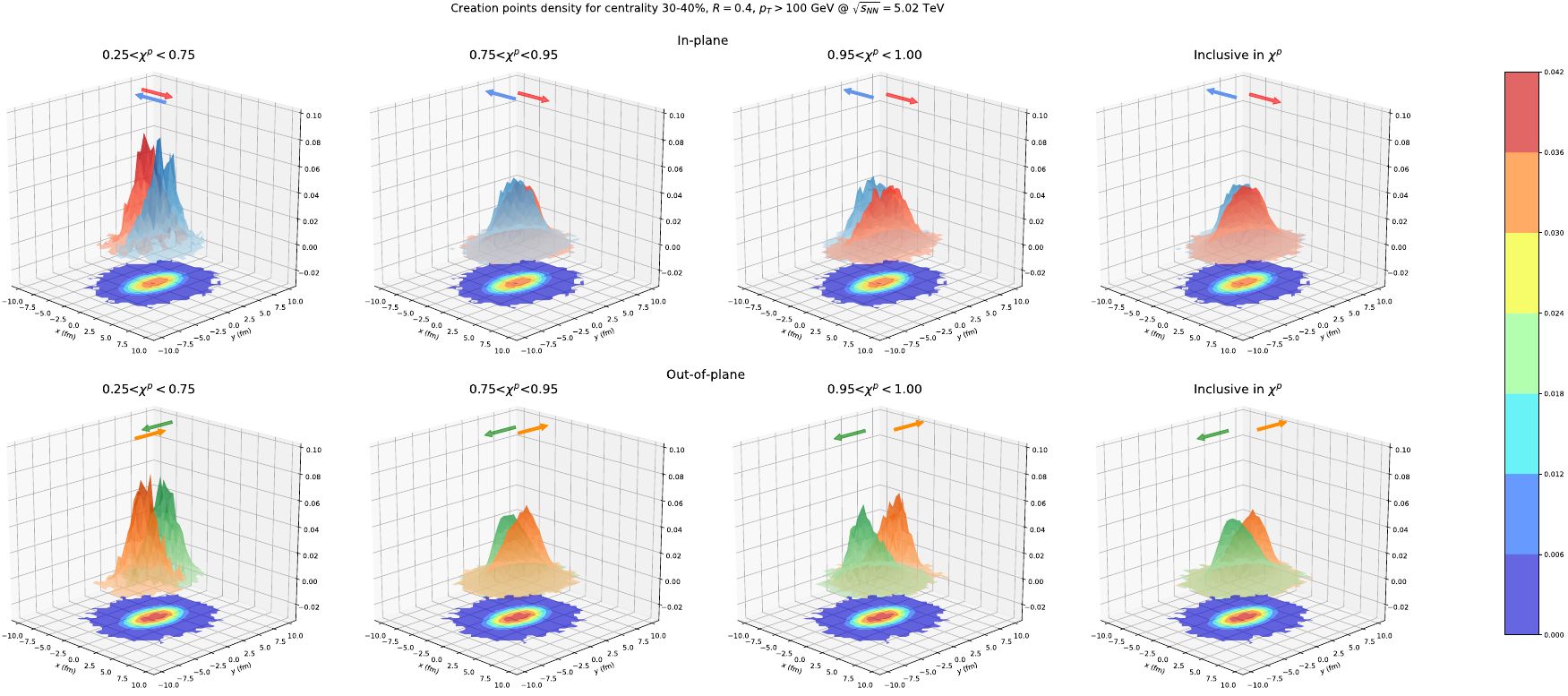}
\caption{Creation point distributions of jets in the transverse plane in 30-40\% centrality for different ranges of the predicted $\chi^p$ in four columns, respectively. The in-plane jets propagating left (blue) and right (red) are shown in the upper row and the out-of-plane jets propagating up (orange) and down (green) are shown in the lower row. The 2-D histogram in the bottom of each plot is the distribution of the inclusive in-plane (upper row) and out-of-plane (lower row) jets in this centrality.}
\label{fig: Tomography_centrality_30_40_in_out_plane_FES_chip} 
\end{figure}
%%%%%%%%%%%%%%%%%%%%%%%%

To improve on locating the creation-point of a dijet pair in the nuclear overlap region more precisely, one could also account for the jet orientation with respect to the event plane of the collision, which is determined by the second azimuthal harmonic of the particle distribution. For demonstration purposes, we will consider jets propagating in-plane, i.e. parallel to the event plane, and out-of-plane, i.e. transverse to the event plane, generated at 30-40\% centrality at $\sqrt{s_{NN}}= 5.02$ TeV for PbPb collisions instead. In Fig.~\ref{fig: Tomography_centrality_30_40_in_out_plane_FES_chip} we show the results of the production-points distribution for around 900,000 jets. In the upper (lower) row we consider the jets that propagate in-plane (out-of-plane), which means they flow approximately along the short (long) axis of the nuclear overlap region. This corresponds to the jets with distinctly positive (negative) $v_2 = \big( p_{{\scriptscriptstyle T},x}^2-p_{{\scriptscriptstyle T},y}^2\big)\big/ \big( p_{{\scriptscriptstyle T},x}^2+p_{{\scriptscriptstyle T},y}^2\big)$. In the bottom of each sub-figure in the upper (lower) row we also show the production-point distribution of the in-plane (out-of-plane) jet inclusive in $\chi^p$. 
We can further select jets according to its propagation direction: either left (in blue) or right (in red) for the in-plane jets, and either up (in orange) or down (in green) for the out-of-plane jets. The histograms in each of fourth column show the results inclusive in $\chi^p$ and corresponds to the production-point distributions if we had no clue to the degree of jet energy loss. One can see some degree of separation, but the overlap is still quite large. This situation improves radically with the help of our knowledge of predicted $\chi^p$. The histograms in each of the first three columns display the production-point distributions for jets belonging to different quenching classes. 

The third column of Fig.~\ref{fig: Tomography_centrality_30_40_in_out_plane_FES_chip} shows results for little quenched jets, with $0.95<\chi^p<1$. To belong to this class, jets have to have traversed merely a short length through the QGP. Focusing on the out-of-plane jets first in the lower row, one can see that the production points of jets propagating upwards are predominantly localized in the \textit{upper} hemisphere of the overlap region (and vice versa for jets propagating downwards). This reasoning also applies reversely for jets belonging to the very quenched class, with $0.25<\chi^p<0.75$, displayed in the first column. One can see that these very quenched jets propagating upwards have had to traverse a long length in the QGP, or analogously through a hot region. Consequently their production points will be predominantly localized in the \textit{lower} hemisphere instead (and vice versa for jets propagating downwards). As the bridge between the little quenched and very quenched jets, the ones with $0.75<\chi^p<0.95$ shows the notably overlapping transition region in second column. Obviously, similar arguments could also apply for the in-plane jets displayed in the upper row of Fig.~\ref{fig: Tomography_centrality_30_40_in_out_plane_FES_chip}.

\section{Summary}
In this proceedings, we review the power of our novel deep learning techniques to locate \textcolor{black}{with precision} the jet creation point in the transverse plane, by selecting jets according to jet energy loss $\chi$, width $R_g$ and orientation, which constitutes a significant step towards the exploitation of tomographic power of energetic jets. 
% To move forward, it will be necessary to study the generalizability of our novel framework by testing the prediction performance of $\chi$ when applied to different energy loss models. 
In future work, the tomographic power is expected to be further improved by considering the interplay between the jet and the local properties of the medium, e.g., the local hydrodynamic flow \cite{Armesto:2004vz,Sadofyev:2021ohn}
% \cite{Yan:2017rku,Tachibana:2020mtb,Casalderrey-Solana:2020rsj}
or spatial-temporal gradients ~\cite{He:2020iow,PhysRevLett.127.082301,Sadofyev:2021ohn}, which determine preferred directions and deformed radiation spectra for the soft emissions from the jet.
%are ingredients that are still missing from the current version of the model used in this work. This extra layer of information, which improves per se the tomographic power of quenched jets, 
A direct extraction of the traversed length of energetic jets in the QGP is highly desirable to push forward this series of studies. 

\acknowledgments
This work is supported by the Trond Mohn Foundation under Grant No. BFS2018REK01 and the University of Bergen. Y. D. thanks the support from the Norwegian e-infrastructure UNINETT Sigma2 for the data storage and HPC resources with Project Nos. NS9753K and NN9753K. D.P. has received funding from the European Union’s Horizon 2020 research and innovation program under the Marie Skłodowska-Curie grant agreement No. 754496.
% \begin{thebibliography}{99}
% \bibitem{...}
% ....

% \end{thebibliography}

\bibliographystyle{apsrev4-1}
\bibliography{duyl}

\end{document}